# AN OPTIMUM ENERGY CONSUMPTION HYBRID ALGORITHM FOR XLN STRATEGIC DESIGN IN WSN'S

Md. Khaja Mohiddin[1] and V. B. S. Srilatha Indira Dutt[2]

[1]Research Scholar, Department of ECE, GITAM (Deemed to be University), Visakhapatnam, Andhra Pradesh, India
[2]Associate Professor, Department of ECE, GITAM (Deemed to be University), Visakhapatnam, Andhra Pradesh, India

***ABSTRACT***

*In this paper, X-Layer protocol is originated which executes mobility error prediction (MEP) algorithm to calculate the remaining energy level of each node. This X-Layer protocol structure employs the mobility aware protocol that senses the mobility concerned to each node with the utilization of Ad-hoc On-Demand Distance Vector (AODV), which shares the information or data specific to the distance among individual nodes. With the help of this theory, the neighbour list will be updated only to those nodes which are mobile resulting in less energy consumption when compared to all (static/mobile) other nodes in the network. Apart from the MEP algorithm, clustering head (CH) election algorithm has also been specified to identify the relevant clusters whether they exists within the network region or not. Also clustering multi-hop routing (CMHR) algorithm was implemented in which the node can identify the cluster to which it belongs depending upon the distance from each cluster surrounding the node. Finally comprising the AODV routing protocol with the Two-Ray Ground method, we implement X-Layer protocol structure by considering MAC protocol in accordance to IEEE 802.15.4 to obtain the best results in energy consumption and also by reducing the energy wastage with respect to each node. The effective results had been illustrated through Network Simulator-II platform.*

## 1. INTRODUCTION

IEEE 802.15.4 standard has its requirements in Medium Access Layer as well as in Physical Layer. It upholds the network topologies based on mesh, cluster, tree & star. In a star topology, the nodes cannot communicate directly without passing through the data collector node through peer-to-peer topology concept, the node itself communicates with irrespective of passing through the sink node which is implemented to multiple network strategies. This standard operates on two different nodes: One is Beacon mode that is Slotted CSMA/CA and the other is Non-Beacon Mode which is Non-Slotted CSMA/CA. Mainly they are dual things that are necessarily to be determined. Out of which, the primary is how to stabilize the energy of the nodes & the second is how to acquire the energy efficiency within the LR-WANs. Universally, all the protocols had to oversee the energy related issues as mentioned below:

1. **Idle Listening**: This state resembles that no transmission is done throughout the simulation process.
2. **Over Hearing**: This state resembles that transceiver is actively listening to the message without worthwhile.
3. **Collisions**: This state occurs when multiple transmissions are done in such a way that the acknowledgment of any one of the message fails.
4. **Over Emitting**: This state resembles the packets that are transmitted when the destination node is not prepared to accept them.
5. **Signaling Overhead**: This state resembles to the energy based on acknowledgments & synchronization of the packets.

The recent advancements in the field with concern to the above aspects have observed a rapid growth in MEMS, low power consumption, huge digital integrated circuits, small scale energy industries, and radio technologies with less power, cost effective and multi-functionality WSNs, which can analyze the changes, occurred in the environment [30]. These devices are integrated with a mini size battery, a mini microprocessor, an aerial, and an array of transducers which converts one form of energy to another, which is used to gain the information that responds to the changes occurred, in the environment of the sensor node [7]. The necessity of these type of device in the WSNs has motivated the emerging research in the recent decades relevant to the potential of association among sensors in data collection and operation, which leads to the era of WSNs [22-23].

The utmost originating prototype IEEE 802.15.4 which has been employed along with the X-Layer model is a standard arising intelligence medium access control protocol which is widely opted for Low-Rate WAN's also it is convenient for mobile sensor networks localization [1]. It is accomplished as a medium of intercommunication linking (MAC) & (PHY) layer [37]. MAC & PHY routing in a zig-zag style along with which it is a structure for secured & data assembling of efficient energy. AODV performs better as compared to DSR (Dynamic Source Routing) protocol with respect to few parameters like high packet delivery ratio, high system throughput, minimum energy consumption as it (AODV) is faster at efficient data circulation [18]. Two-Ray Ground method is taken into consideration in this paper as a propagation model in contrast with the shadowing method as it is having genuine performance in some of the parameters like energy consumption, end-to-end delay & system throughput [19-20].

Table 1 Different Approaches & Methods of IEEE 802.15.4 based MAC Layer

| Types | Definition |
|---|---|
| Parameter Tuning Method | Tuning of super-frame parameters to enhance the performance without rectifying the level & specification of IEEE 802.15.4. It is entirely dependent upon the application & its performance is based upon its parameter's value. |
| Cross-Layer Method | Provides key solutions depending upon the influence of various layers within the protocol, but leads in enhancement in latency. |
| 802.11 Method | Relocates the key solutions that had been projected by it to 802.15.4 environment. It has a capability of re-utilizing the experimented technology where power consumption has no priority in this method. |
| Priority Method | Improves the Quality of Service assistance so that all the nodes along with its traffic is been given the priority where power consumption has no importance. |
| Duty Cycle Method | Manages the active frames to obtain maximum power preservation with minimum manipulations. |
| Back-Off Method | Provides dynamic assistance to various topologies which requires no hardware up-gradation & may lead to maximum manipulation of the standard. |
| QoS Method | Better assistance for applications based on time-sensitivity. |
| Hidden Term Resolution | Drastic collisions are being minimized with less packet re-transmissions. |

## 2. Related Work

According to the existing scenario of WSNs, it is exclusively mandatory to enhance the network specification performance in which XLN has a major role [3]. Adjustment of the transmission power certainly outperforms the deduction of the requirement of energy consumption [5]. Apart from the above solution, there is the possibility of another approach where the parameter 'ED' has being estimated to overcome the drawback of control overhead, where E represents the energy and D represents for the degree. In this paper, the results are compared with EQSR, X-layer as well as Enhanced Model in which the enhanced model is actually the proposed and performs better in terms of efficiency evaluation [6]. EQSR mainly focuses to identify the best linking route. The crucial drawback of the X-layer method used in this paper is that it sustains from broadcast overhead, if the node has maximum mobility which results in huge consumption of energy during the establishment of the link discovery process. In, all the 3 methods were being compared and it was observed that energy consumption per packet is less in the proposed(ED) model along with the increase in network lifetime. Also, the channel occupation as well as energy consumption is minimized which also results in minimization of the in the delay parameter [3-4].

In this paper, the authors implemented a mobility based clustering algorithm for WSN with few mobile nodes. According to this new algorithm, a sensor node elects itself as a cluster-head based on its remaining energy level available with it and mobility parameter. Whereas, a non-cluster head, focuses on its link stability with CH during clustering within the estimated duration of time. The Individual time slot for data transmission is assigned to the non-cluster head node in increasing order in a TDMA scheduled within the given time. During the static-state, a sensor node sends its sensed data in TDMA pattern as slots and transmits the REQ message to join under a new cluster also it avoids the loss packets to stable its connection with its cluster head [21, 26].

There is an innovative approach for routing as well as for clustering with respect to the grid view in WSN. Knowing the area of the deployment & transmission range, the grid size is being estimated and thereafter the CH is selected depending upon the nearest available mid-point of the grid. Here the proposed algorithm in this paper outperforms better when compared to the existing methods like GCMRA, LPGCRA. In this GCCR which represents Grid Based Clustering & Combinational Routing is established where CH is assigned knowing the residual energy and is more eligible for hybrid WSNs [5].

As the mobile nodes continuously change the network coordinator at regular intervals, it has been observed that in active condition of the beacon, IEEE 802.15.4 can't sustain the inter-connectivity for that particular nodes which may lead to a major issue in the association of the node's with the network coordinators [36]. Due to the various nodes with different mobility's within the network, it may lead to improper performance & malfunctioning in the synchronization with the network coordinators. When the respective network node exits out of the concerned cluster range, then it has to directly associate in looking out the new coordinator/cluster without providing the loneliness notification so that the energy can be saved to a certain extent [6-7].

As far as the on-demand routing protocols is concerned in wireless sensor networks, they are majorly three types namely DSR, DSDV & AODV. After analyzing the simulation results of various specifications such as packet loss, energy consumption & average delay, the AODV's energy consumption is around 18.07% less when compared to DSDV [38-39].

Wireless Sensor Networks has its own importance when IEEE 802.15.4/Zigbee comes into existence in-terms of development or enhancement where a cross layer model has been addressed to choose such a network coordinator which depreciates the energy consumption requirement of the nodes, reduces repetition transmission of the same packet and ease of association of the node with other network coordinators [40].

To achieve the congestion control, MAC & efficient routing mechanism, cross layer protocol has been implemented where it consists of such a features which makes the comparison of functionalities easier with the help of thresholds. It is first protocol which has layer-to-layer communication [13]. It is completely based on the architecture of layered protocol in terms of providing good performance & critical implementation [14].

The XLP protocol with respect to the Code Division Multiple Access Ad-hoc networks is concerned, the energy optimization can be done more efficiently if the route request messages could be stored during the process of route establishment when packets have been transmitted more in number due to which the number of hops also increased [35]. Corresponding to this issue, a modified XLP protocol is been combined with AODV protocol to provide more ease in extracting the shortest path towards the destinations [32-33].

## 3. Features Of IEEE 802.15.4

Multiple sensors occupies a small vacancy which states a word called "wireless", thereafter resulting in low-power which in-turn implies finite distance. With respect to this, all the nodes should be in a self-organized manner for which low cost approaches can be implemented to achieve the concerned target. Wireless approaches have following benefits like: connectors are not required, secure & reliable connectivity, increasing inflexibility of sharing resources, mobility & effortless installation [8]. Each & every node has its own bounded range to which the message has to be transferred. But if at the range is more than the desirable range then it takes the help of the other nodes so that the message can be reached to the destination safely, this mode of communication is known as "Multi-hop Communication". In this, the network itself changes the topology in the wireless environment [15]. 802.11 relates to the Wireless LAN concept where it is used in a centralized way in the form of Embedded Sensors within a block (or) building (or) office etc which costs very huge for its implementation. Whereas 802.15.1 commonly known as Bluetooth can be used as Wireless PAN which is the combination of both Video as well as WLAN as shown in Figure (1) where the cables are being replaced by a protocol which costs moderate. Thereafter, 802.15.4 commonly referred to Zigbee used as a sensor as well as actuator devices for the industrial & commercial use with cost effective property & operates for real-time applications [25, 31].

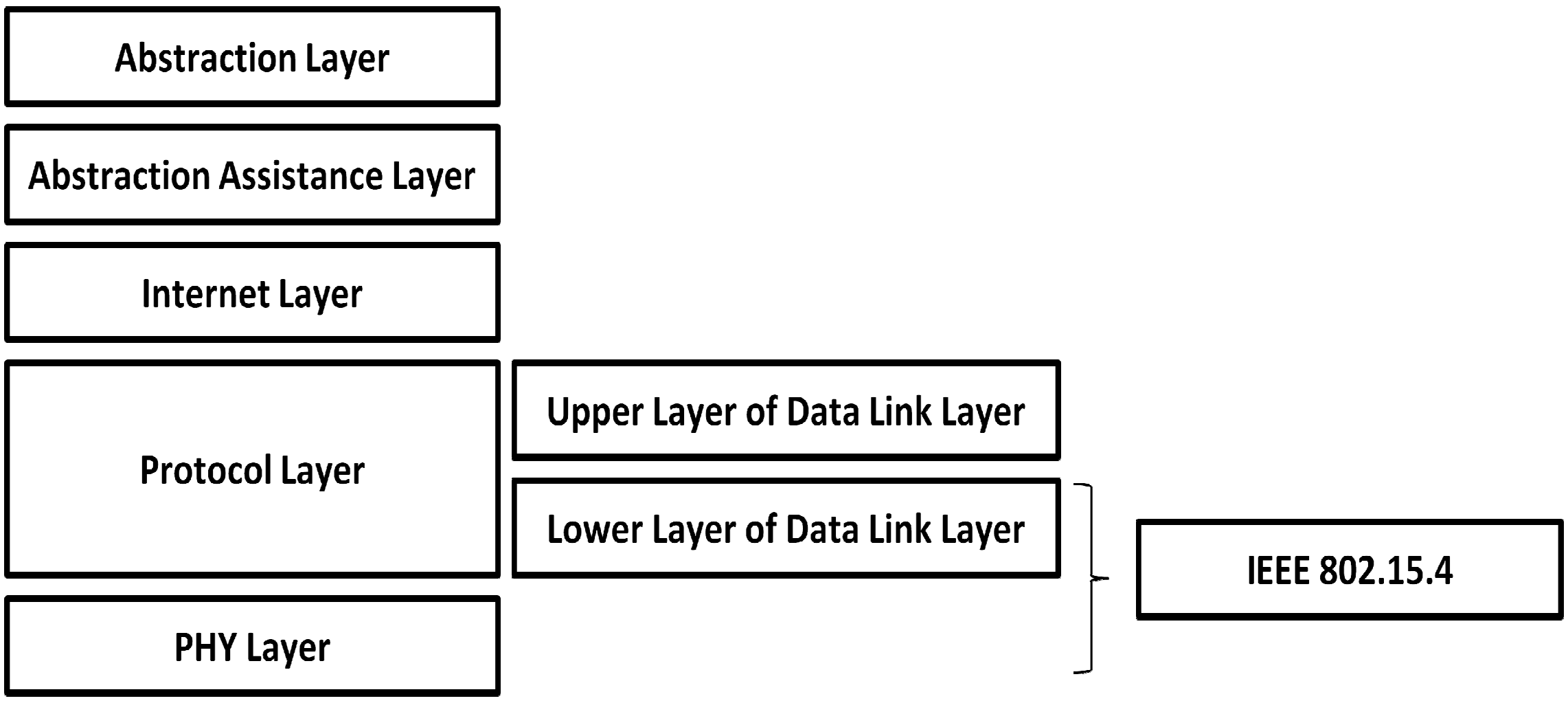

Figure 1 Wireless Networking Protocol Stack Model

It operates with the data rates of 20 Kb/sec, 40 Kb/sec & 250 Kb/sec with dynamic device addressing. It also has the special characteristics like less power consumption, fully handshake protocols for transmission etc [10]. It has the frequency bands of operation as follows with respect to Figure (2):

1. 2.4 GHz ISM-Bandwidth “16” Channels
2. 915 MHz ISM-Bandwidth “10” Channels
3. 868 MHz with “01” Channel

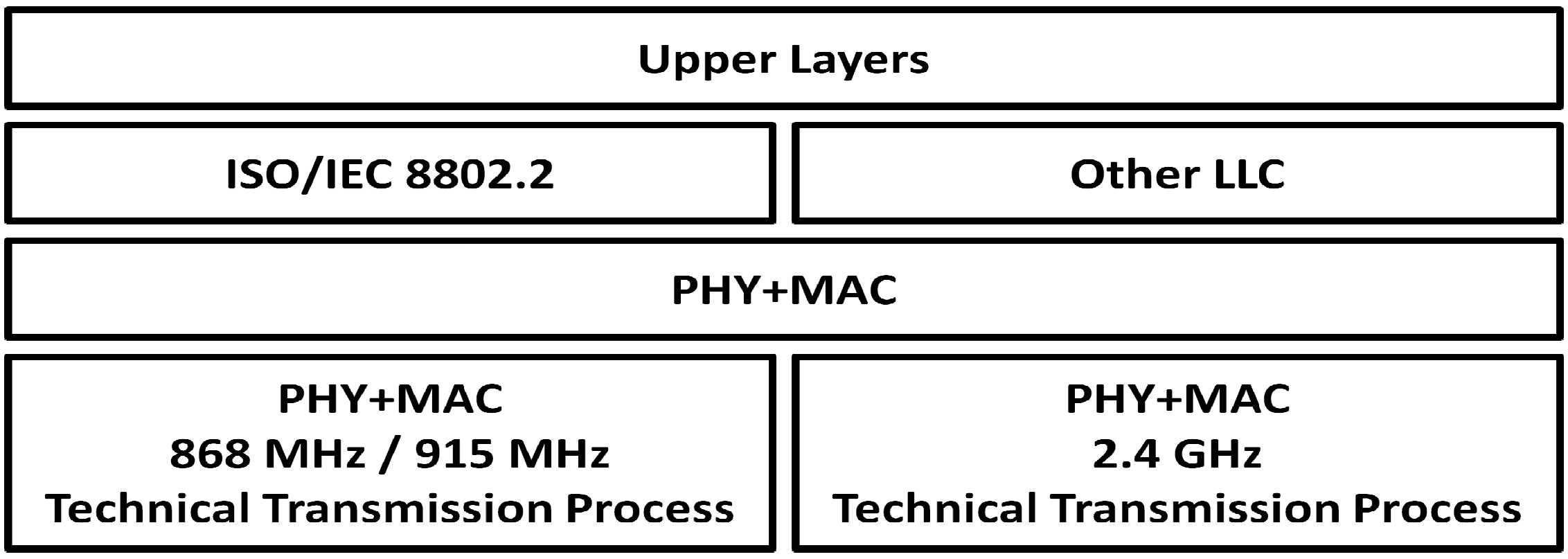

Figure 2 Protocol Architecture

IEEE 802.15.4 has two device type classes namely:

1. Full-Function Device
2. Reduced-Function Device

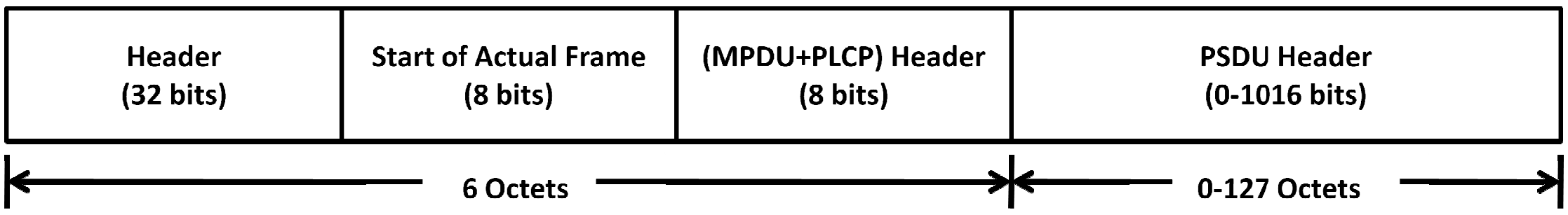

Figure 3 Packet Structure

The FFD is capable to access in any type of topology along with the monitoring of Personal Area Network Coordinator Capabilities to interact with any other devices too. Whereas RFD is limited to star topology & does not has the capability to become a network coordinator with easy implementation to interact with only one network coordinator. FFD (or) RFD contains the IEEE 802.15.4 in which it manages the interface of the both MAC sub-layer as well as PHY layer to the wireless medium [29]. Frame Composition of IEEE 802.15.4 MAC is of 4 types like Data, Beacon, MAC Command & Acknowledgement Frames whereas the Super Frame Composition of IEEE 802.15.4 MAC also contains few types as follow:

1. **Network Beacon**: This beacon is being transmitted by the Personal Area Network Coordinator which contains its information, frame structure along with the notifications of the awaiting node information.
2. **Beacon Extension Period**: This period is the space reserved for the awaiting notifications.
3. **Contention Period**: This can be monitored by any node using Carrier Sense Multiple Access-Collision Avoidance.
4. **Guaranteed Time Slot**: This slot is reserved for such nodes which require mandatory bandwidth.

IEEE 802.15.4 MAC has few types of traffic relevant features namely: Periodic Data (sensor applications based on application defined), Intermittent Data (light switch based on external stimulus defined) & Repetitive Low Latency Data (based on time slot allocation applications as elaborated in Figure (3).

IEEE 802.15.4 MAC has various applications as follows:

1. Residential Heartbeat System
2. Residential Home Awareness System by utilizing the sensors like temperature, water, power etc.
3. Wireless Lighting Control
4. Energy Conservation for industrial use based on motor/system efficiency along with safe sensing applications
5. Energy Sensing Applications
6. LRWPAN Field Test Site
7. Energy Savings based on closed loop
8. Securing Energy on Infrastructure Capabilities
9. Prediction Based System

## 4. METHODOLOGIES USED IN THE PROPOSED SYSTEM MODEL

### 4.1. Mobility Error Prediction (MEP) Method

In this MEP method, the mobility of each and every node is being estimated depending on its deviation from it's initial location. It is not necessary that every node should be mobile, some of the nodes may be mobile and others may be static [34]. So, to identify the node's status, the mobile aware protocol is being executed in the form of calculating it's deviation in X-Axis & Y-Axis from it’s actual position. From existing, we change the location discovery process by the adaptive beaconing model which integrates beaconing only after moving from current to new location with distance threshold (For Example: >5m). Also, this process change the energy consumptions depending on distance they transmit, so we change the energy formula by following methods:-

$$\boldsymbol{Remaining\ Energy}\ (\boldsymbol{Node}) = \sum \{T_{xn}(e) \div D + R_{xn}(e) + Idle(e) + Sleep(e)\} \quad (1)$$

Where,

| | | |
|---|---|---|
| $\boldsymbol{T_{xn}(e)}$ | = | Energy consumed during transmission of all packets as per distance |
| $\boldsymbol{R_{xn}(e)}$ | = | Energy consumed during reception of all packets as per distance |
| $\boldsymbol{Idle(e)}$ | = | Energy consumed during idle node waiting for request time |
| $\boldsymbol{Sleep(e)}$ | = | Energy consumed during sleep mode (after long idle mode, the node goes to sleep mode) |

### 4.2. Cluster Head Selection

Cluster Head (CH) Selection should be done at regular intervals so as to identify the remaining clusters existing within the same region or not. Apart from the above, all nodes also need to select their corresponding cluster by sending the signals in form of beacons.

### 4.3. Clustering Multi-Hop Routing Method

If at all the cluster head (CH) is at a certain distance from the base station, then it needs to select multiple clusters to occupy the location as well as position nearby the base station as explained in Figure (4). The simulation parameters along with the considerations of various protocols and models are being mentioned in Table 2.

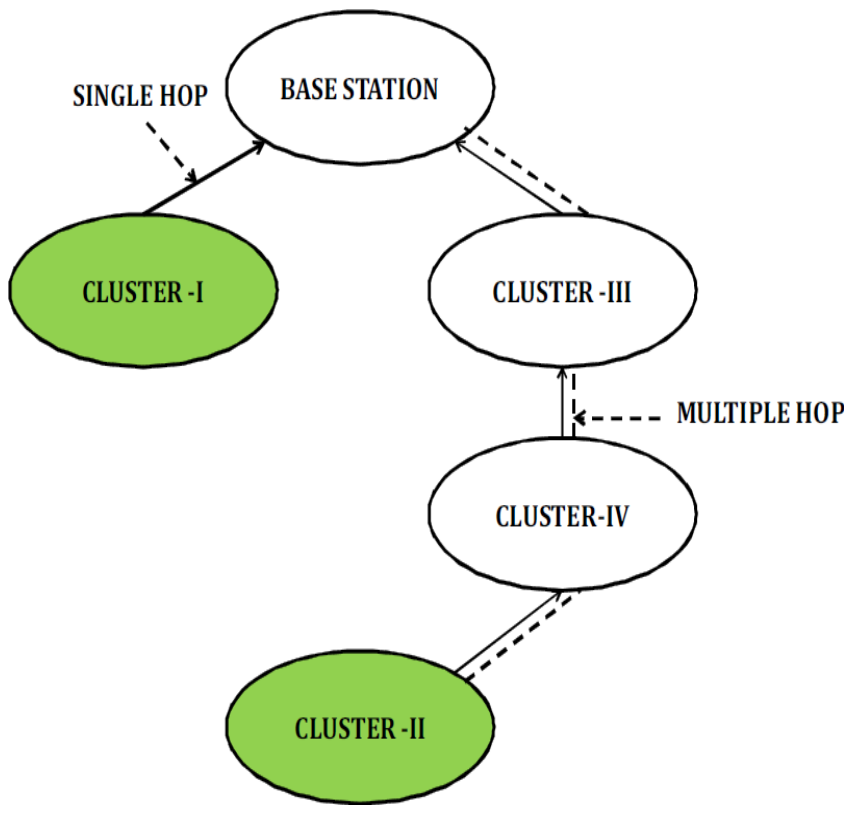

Figure 4 Clustering Multi-Hop Routing

Table 2 Simulation Parameters & its Specifications used for analysis.

| **Simulation Parameters** | **Specifications** |
|---|---|
| No. of Sensor Nodes | 10-100 |
| Initial Energy | 100 Joules |
| Network Size | 250 x 250 |
| Sink Location | Centre of the Region |
| Mobility | 0 m/s to 5 m/s |
| Transmission Range | 35 meters |
| Simulation Time | 100 seconds |
| Radio Propagation | Two-Ray Ground Model |
| Transport Protocol | User Datagram Protocol |
| Application Protocol | Constant Bit Rate (QoS) |
| MAC Protocol | IEEE 802.15.4 |
| Packet Size | 256 Bytes |
| Queue Length | 100 Packets |
| Routing Protocol | Ad-hoc On Demand Distance Vector |
| | Mobility Aware Secure Routing Protocol |

## 5. PROPOSED SYSTEM MODEL

If the movement of the nodes is confined to a few nodes likewise sink nodes, and then the stable nodes can be mitigated in terms of routing link originated towards the node till it reaches the destination. The base station nodes can move around through stable nodes & accumulate the information sensed by the source nodes via beacon. The collector nodes which are mobile may also improve the network link connectivity through reducing the bottleneck problem that may arise during the network traffic overflow. In the existing methodology, it is being observed that the location discovery process is implemented in the application layer of the corresponding architecture through beaconing process to track the position of the node at regular intervals. This may also result in improvement of the consumption of energy in terms of packet delivery ratio.

### 5.1. Design Assumptions:-

1. The system model is analogous.
2. All the source nodes were mobile so as to overcome the drawback of non-beaconing.
3. A stable data collector node is being deployed at the center of the cluster.
4. The deployment scenario is flat.
5. Vicinity range of each other what is within the Line of Sight for successful transmission.

In this paper, it is has proposed that the XLN (cross-layer network operation) model [2] whose initialization process begins with the broadcasting of the NB discovery request to originate the NB information assembling and store it in the NB-List. As soon as the above mentioned process is initiated, the position, location & detailed information of the node are shared either by GPS device or by other means of communication so that it can gather and estimate the location of the individual nodes. Then the correspondent node starts sending the route-request packets to construct the route link towards the destination node so that it is entirely feasible to prefer NB list from the N-layer within the D/L layer.

In Figure (5), the flow chart of the proposed system has been discussed originating from the initial deployment process where the nodes take their initial positions, thereafter the positions are being updated via adaptive beaconing process through which the neighbor discovery process can be initialized for updation of the neighbor list. After this, the route discovery process will be initiated by sending the Route Request Packets (RREQ), if it receives the acknowledgment then it calculates the energy as well as link metrics based on which an optimal path is being chosen. After the completion of this whole process, it initiates the data communication, but if in between the link changes due to the mobility then once again the updation of the neighbor list is required.

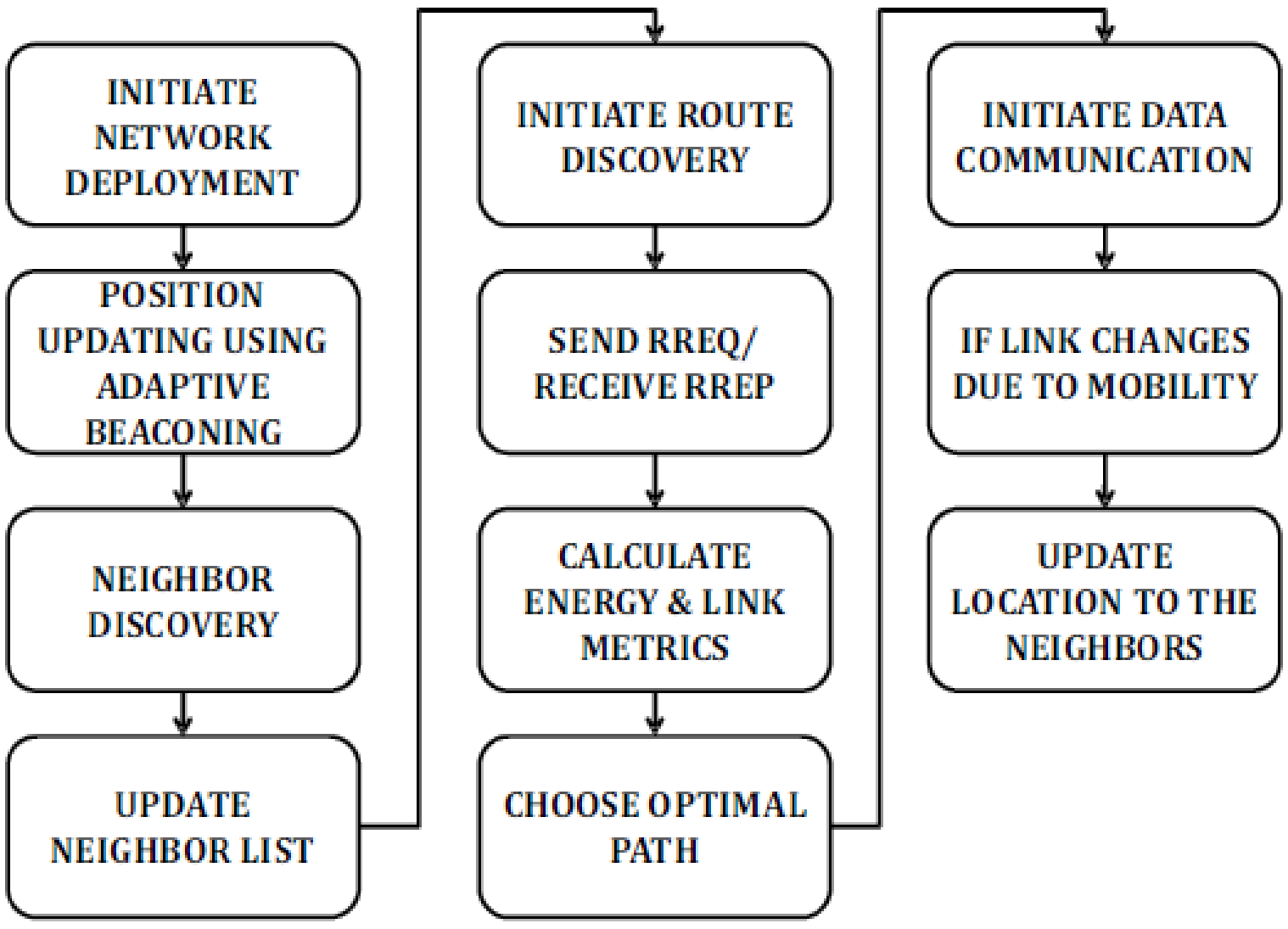

Figure 5 Proposed System Flow Chart

### 5.2. Comparison of Proposed Methodology with the Existing System:-

As per the existing system is concerned, the cross layer approach has been implemented to compensate the drawbacks occurred in the energy consumption [1]. Apart from this, the WSN also suffers from huge link failures [12]. To overcome the above problems in the existing system, an innovative approach is been proposed in this paper combining 3 different algorithms to minimize the issues occurring in the existing model. First algorithm identifies which are mobile and also considers the static nodes for future communication which minimizes the energy consumption. Second elects a proper cluster head so as the node can stay within the same cluster for longer duration to avoid link failures. And the third algorithm implements a multi-hop clustering along with the above two algorithms. By contrasting all the 3 algorithms, we can observe maximum changes in the energy consumption with less packet loss and deserved system throughput.

## 6. ALGORITHM IMPLEMENTED

### 6.1. Algorithm 1: Forwarding Node Mobility Error Prediction for Next Possible Relay

Input:

$\boldsymbol{n_i}$ = Nodes (1, 2, 3……i)

$\boldsymbol{x_i}$ = X-Coordinate with respect to the $i^{th}$ node

$\boldsymbol{y_i}$ = Y-Coordinate with respect to the $i^{th}$ node
$\boldsymbol{t_i}$ = Time with respect to the $i^{th}$ node

Output:
Mobile Deviation

**Step 1** : Mean Distance in X-Axis by $i^{th}$ node
**Step 2** : Initiate Neighbour

$$\mathbf{P}(\boldsymbol{\mu}_\mathbf{x}) = \frac{x_1t_1 + x_2t_2 + x_3t_3 \ldots\ldots\ldots\ldots\ldots . x_kt_k}{k} \quad (2)$$

Here it calculates the probability mean distance moved by the node in time 1, 2, 3.....k in X-Axis

**Step 3** : Mean Distance in Y-Axis by $i^{th}$ node
**Step 4** : Initiate Neighbor Discovery Process & Share the Location

$$\mathbf{P}(\boldsymbol{\mu}_\mathbf{y}) = \frac{y_1t_1 + y_2t_2 + y_3t_3 \ldots\ldots\ldots\ldots\ldots . y_kt_k}{k} \quad (3)$$

Here it calculates the probability mean distance moved by the node in time 1, 2, 3.....k in Y-Axis

**Step 5** : Distance Variations

$$\mathbf{D} = (x_i + P(\mu_x))^2 + (y_i + P\left(\mu_y\right))^2 \quad (4)$$

**Step 6** : When the 'D' is varied for maximum 5m distance, then Source disconnects the link between the neighboring nodes
**Step 7** : Receiving Node Error Reporting
**Step 8** : Calculate remaining energy level by general energy parameters

$$\mathbf{E_c} = (E_{rxn} \times N_{rx}) + (E_{txn} \times N_{rx}) + (P_i \times T_i) + (P_s \times T_s) \quad (5)$$

Where,

$E_{rxn}$ = Energy Consumed during Reception Mode
$E_{txn}$ = Energy Consumed during Transmission Mode
$N_{rx}$ = No. of Packets Delivered (Received) by the Node
$N_{rx}$ = No. of Packets Sent (Transmitted) by the Node
$P_i$ = Power Consumption during Idle Mode
$P_s$ = Power Consumption during Sleep Mode
$T_i$ = Node Spend Time during Idle Mode
$T_s$ = Node Spend Time during Sleep Mode

$$\mathbf{E_r} = \text{Remaining Energy Level} = (\text{Initial Energy}) - E_c \quad (6)$$

**Step 9** : Returning to the Next Possible Relay Based on Remaining Energy Level.

In the section (6.1), the algorithm elaborates that the distance of the node has been calculated as per the mobility parameters in X-Axis & Y-Axis [9]. The mobile nodes start moving from one place to another, if the distance variation from its previous location in more than 5m (For Example) then the nodes starts sharing it's information via adaptive beaconing process through which the base station can also calculate the remaining energy level of each node and can estimate the what range up to which it can utilized.

## 6.2. Algorithm 2: CH Election Procedure

**Cluster Head Election Procedure**

**Step 1** : Each node calculates it waiting time, $W_i^T$ based on its residual energy

$$\mathbf{W_i^T} = \left[1 - \frac{E_i}{E_{max}}\right] T_2 . V_r \quad (7)$$

Where $\mathbf{E_i}$ is Residual energy of $i^{th}$ node & $\mathbf{E_{max}}$ is the extreme initial energy of the relevant nodes in the network, $\mathbf{V_r}$ is a random real parameter value, which is invariably disseminated in the interval [0.9, 1].

**Step 2** : IF $\mathbf{T_1}$ (expires)
Initiate CH election
IF ($\mathbf{N_i}$ not receives any $ADV_{CH}$)
IF ($\mathbf{W_i^T}$ expires)
End $ADV_{CH}$
End
Else (After receiving the $ADV_{CH}$)
$\mathbf{N_i}$ Receives the advertisement from $\mathbf{N_j}$ & maintains the node ID as well as power level
Calculate Dist ($\mathbf{N_i}$,$\mathbf{N_j}$)
Set $\mathbf{N_i}$ non-cluster head node
End

**Step 3** : Calculate Overlying Radius ($\mathbf{N_i}$,$\mathbf{N_j}$)

$$\mathbf{R_{oi}} = [1 - \alpha(d_{max} - D_i) \div (d_{max} - d_{min})]R_{max} \quad (8)$$

Where $\mathbf{d_{max}}$, $\mathbf{d_{min}}$ are maximum as well as minimum distances from the corresponding base station.

$\mathbf{D_i}$ Represents the distance from the base station to the relevant node, '$\boldsymbol{\alpha}$' is the random value [0; 1], $\mathbf{R_{max}}$ is the maximum value of the tolerable competition radius.

**Step 4** : Calculating $\mathbf{R_{oi}}$, the single maximum energy node will choose itself as a CH node.

**Cluster Formation**

**Step 5** : Calculate the Dist $(\boldsymbol{CH}, \boldsymbol{CH_i}) \div \boldsymbol{CH_j}$ Where the CH advertisement is received.
**Step 6** : Send the join cluster message to member nodes (ID, Residual energy)
**Step 7** : CH formulate the node-schedule (NS) list including $\boldsymbol{SCH_{msg}}$ for its members of the cluster.
**Step 8** : After receiving the $\boldsymbol{SCH_{msg}}$ Member nodes get an idea about the time of transfer of data to the CH node.

In the section (6.2), the cluster needs to search & verify that how many nodes exist within the same region for which $\mathbf{W_i^T}$ is being estimated to represent the time taken for the nodes spending in the same region. Further it is required to calculate the residual energy between the error value 0.9 and true value 1. Clustering has to done as it saves the energy in wireless sensor network environment & also it needs to be placed at the centre of the location [11]. Cluster Head Selection is entirely dependent on its energy level whereas the spending time of the node in the same location is based on its position [5].

### 6.3. Algorithm 3: Clustering Multi-Hop Routing

**Flowchart Representation**

**Step 1** : Estimate the total number of cluster heads based on region (or) no. of nodes
**Step 2** : Calculate the grid size of mobile nodes in corresponding region
**Step 3** : For all node placed in region calculate min and max of (x, y)
**Step 4** : For $i = 1$ represents no. of nodes in region
If $\min(x) \leq node.x \leq \max(x)$ and
$\min(y) \leq node.y \leq \max(y)$, then
$CH(i) \ll node.ID$
$D_i(j) \ll$ Distance between midpoint and node location
End if
End for
**Step 5**: Calculate distance of the elected CH with its neighbors

$$\mathbf{D(CH, n)} = \sum_{i=0}^{N} \sqrt[2]{\{CH_i(x) + n_i(x)\}^2 + \{CH_i(y) + n_i(y)\}^2} \tag{9}$$

**Step 6:** Choose the CH which has minimum distance to reach their neighbor

$$CH \leftarrow \min[D(CH, n)] \tag{10}$$

In this section (6.3), the total number of the clusters are been identified for the alignment of the CH within the respective regions [27]. The CH representation may be preferred either by single-hop routing or by multi-hop routing. After which, the grid dimensions are being estimated so as to place all the nodes within the region of X & Y. To calculate the distance of the each node from its cluster, which is to be elected is done by estimating the distance of each cluster with its neighbors by taking the consideration of the node ID as well as its minimum & maximum value of the dimensions in X & Y-Axis by evaluating the equation mentioned the section (6.3)[5].

## 7. Results And Discussion

To estimate the performance evaluation characteristics of the XLN (Cross Layer Network) model, extensive simulation results and its evaluations have been performed using NS-2 platform.

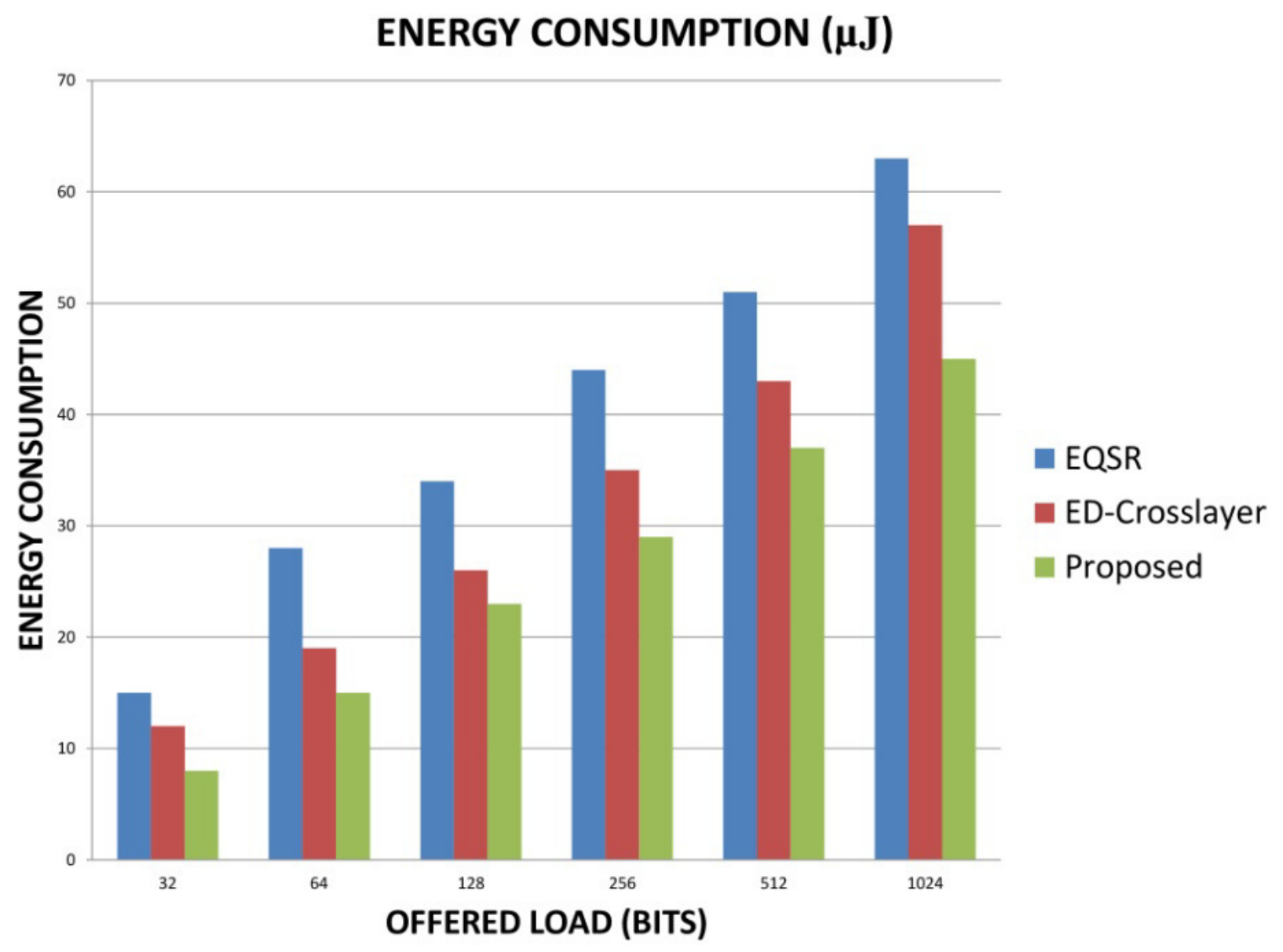

Figure 6 Energy Consumed Per Offered Load in Bits

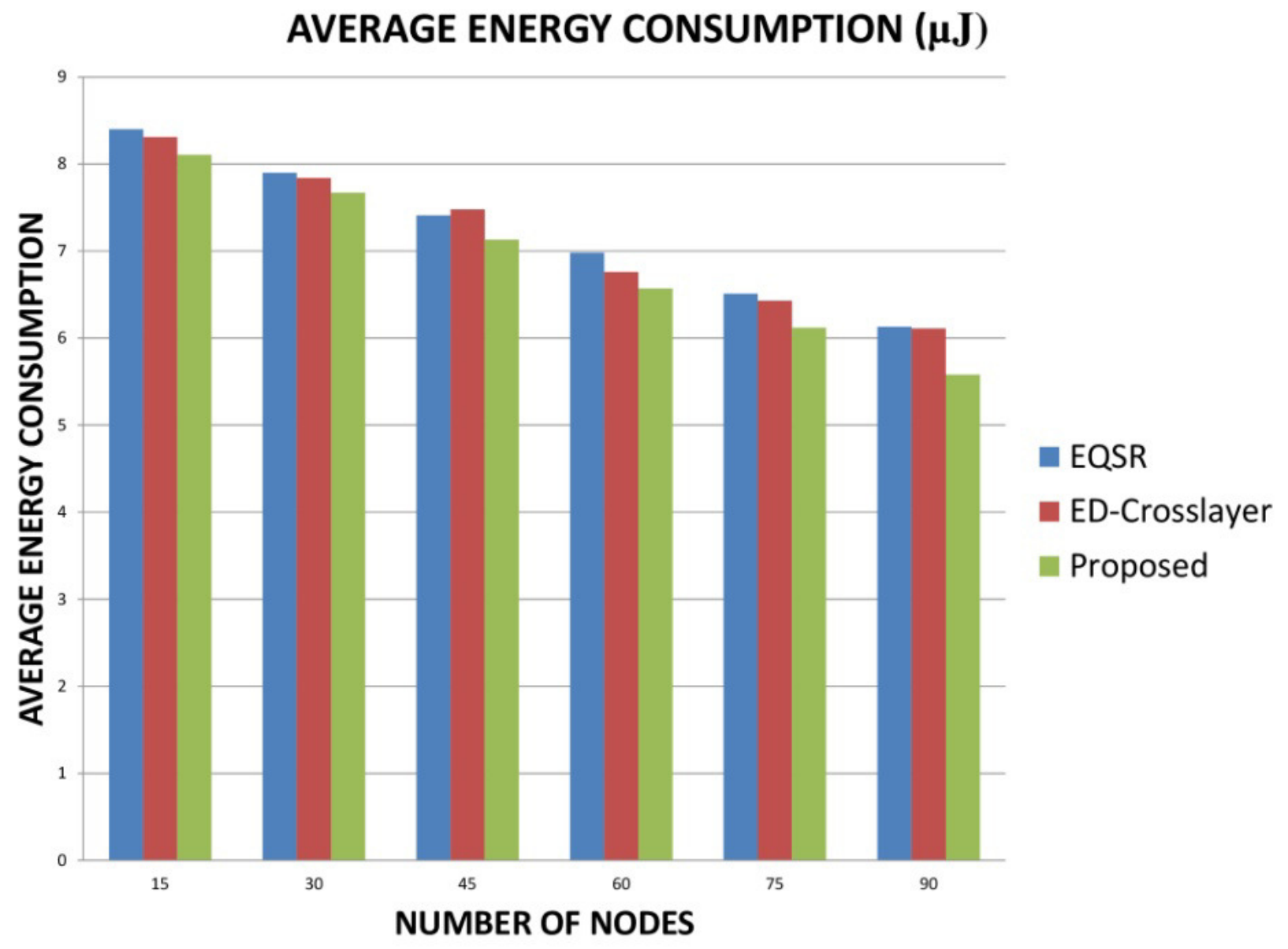

Figure 7 Average Energy Consumed Per Number of Nodes

The proposed approach mainly focuses on the mobility awareness protocol which has been suggested as well as preferred with the help of mobility error prediction model so that the base station can be aware of mobility of the nodes through its adaptive beaconing signal & also assists to know the remaining energy levels of the corresponding node which can be used for further transmissions where the node can be selected with respect to the distance it used for communication. Taking into consideration, the concept of Cluster Head Selection via GCCR algorithm [5], the algorithm 2 has been proposed and thereafter the performance evaluation of this proposed model has been compared with EQSR model as well as ED (Enhanced) of the cross layer model and outperforms better results [4].

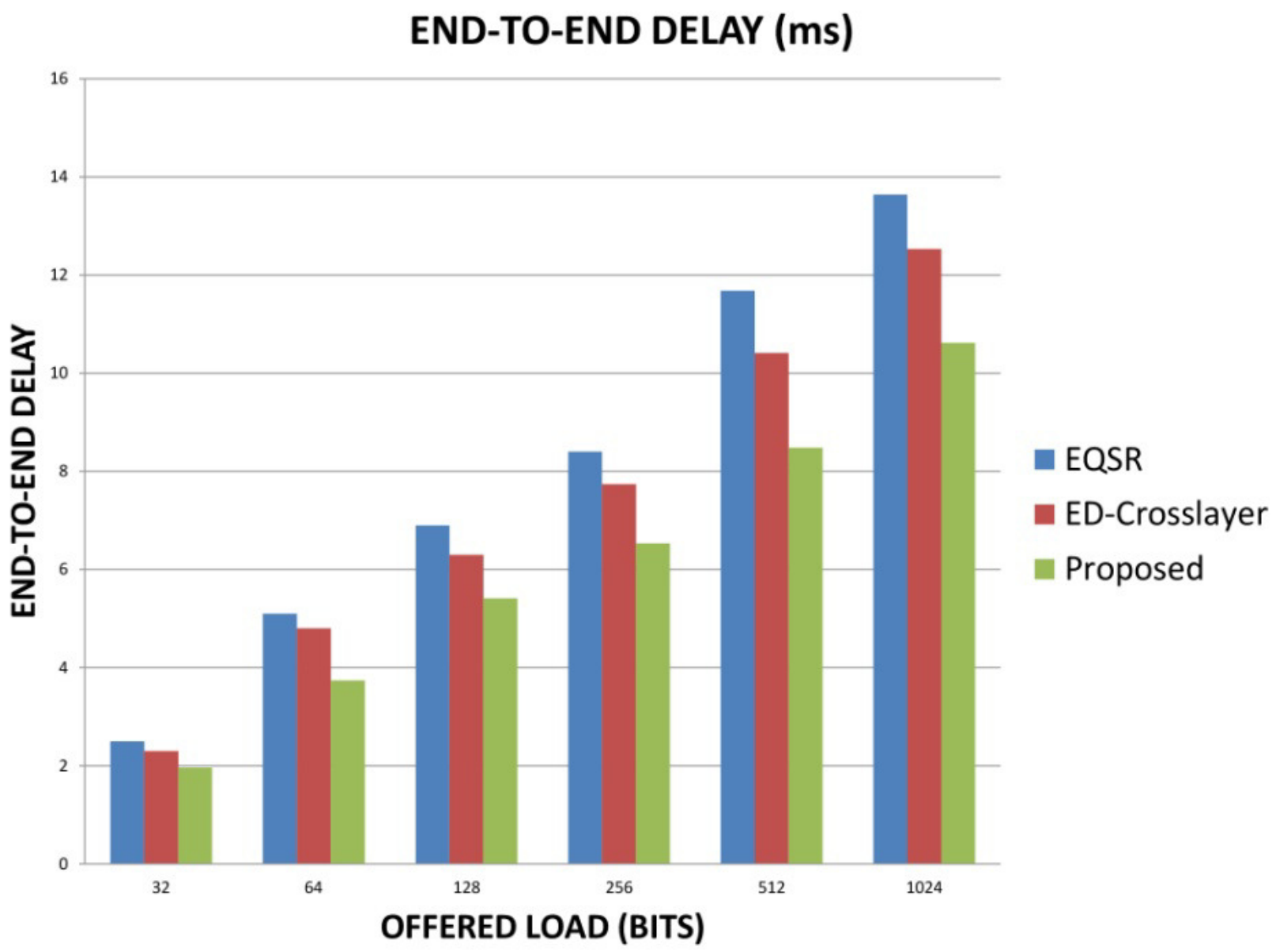

Figure 8 End-to-End Delay Per Offered Load in Bits

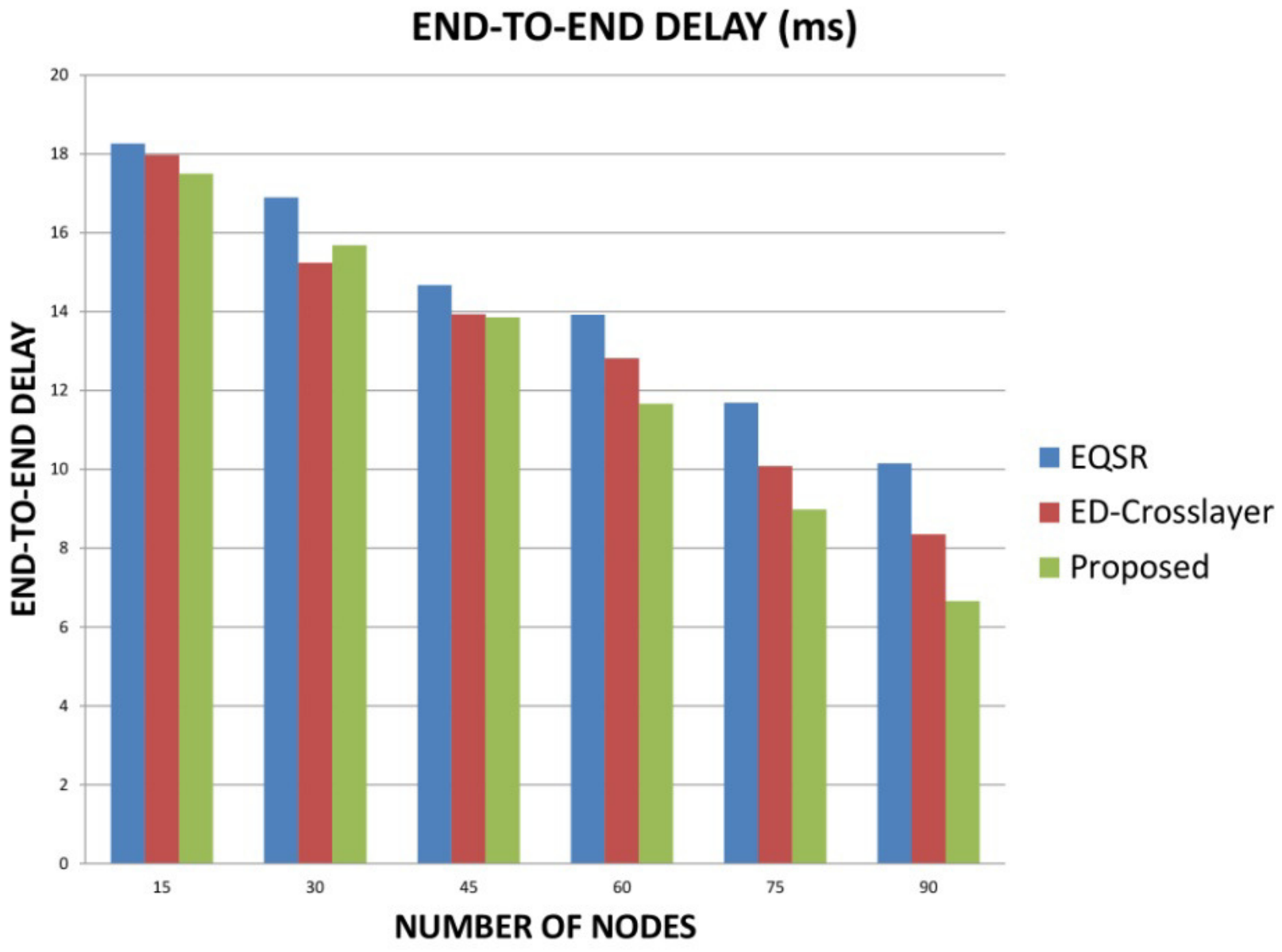

Figure 9 End-to-End Delay Per Number of Nodes

**Energy Consumption Evaluation**: The energy consumed is expressed in micro joules with respect to the offered load (bits) representing the data that has to be transmitted in parallel which is expressed in Figure (6). Whenever the offered load is amplified, energy consumption has also been increased. The energy consumed is also expressed in micro joules with respect to the number of nodes which is shown is Figure (7). As the no. of nodes has been improved, the average energy consumption gets reduced as they will availability of lot of mobile nodes with different energy levels to choose the node with shortest path with high energy levels.

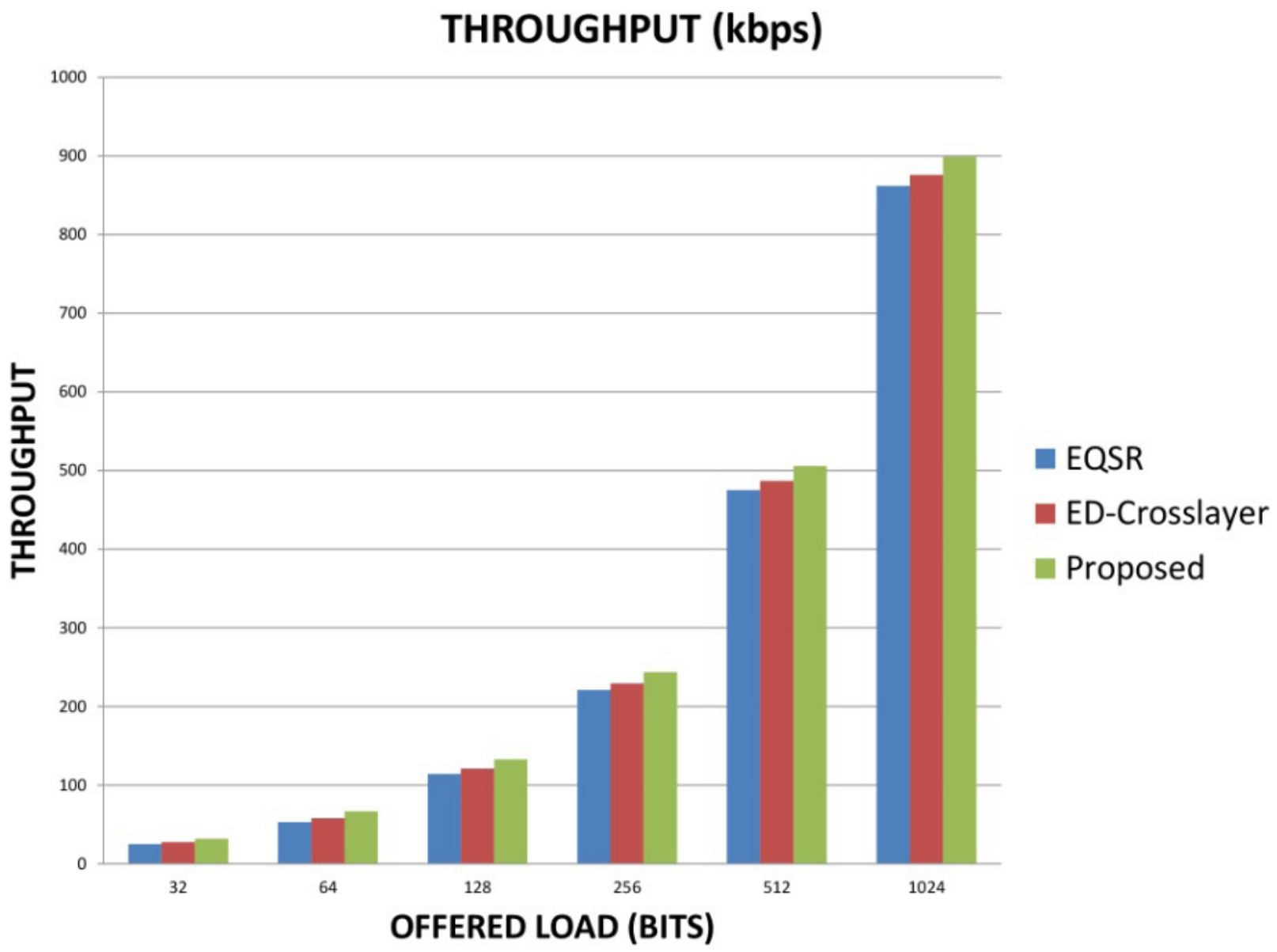

Figure 10 System Throughput Per Offered Load in Bits

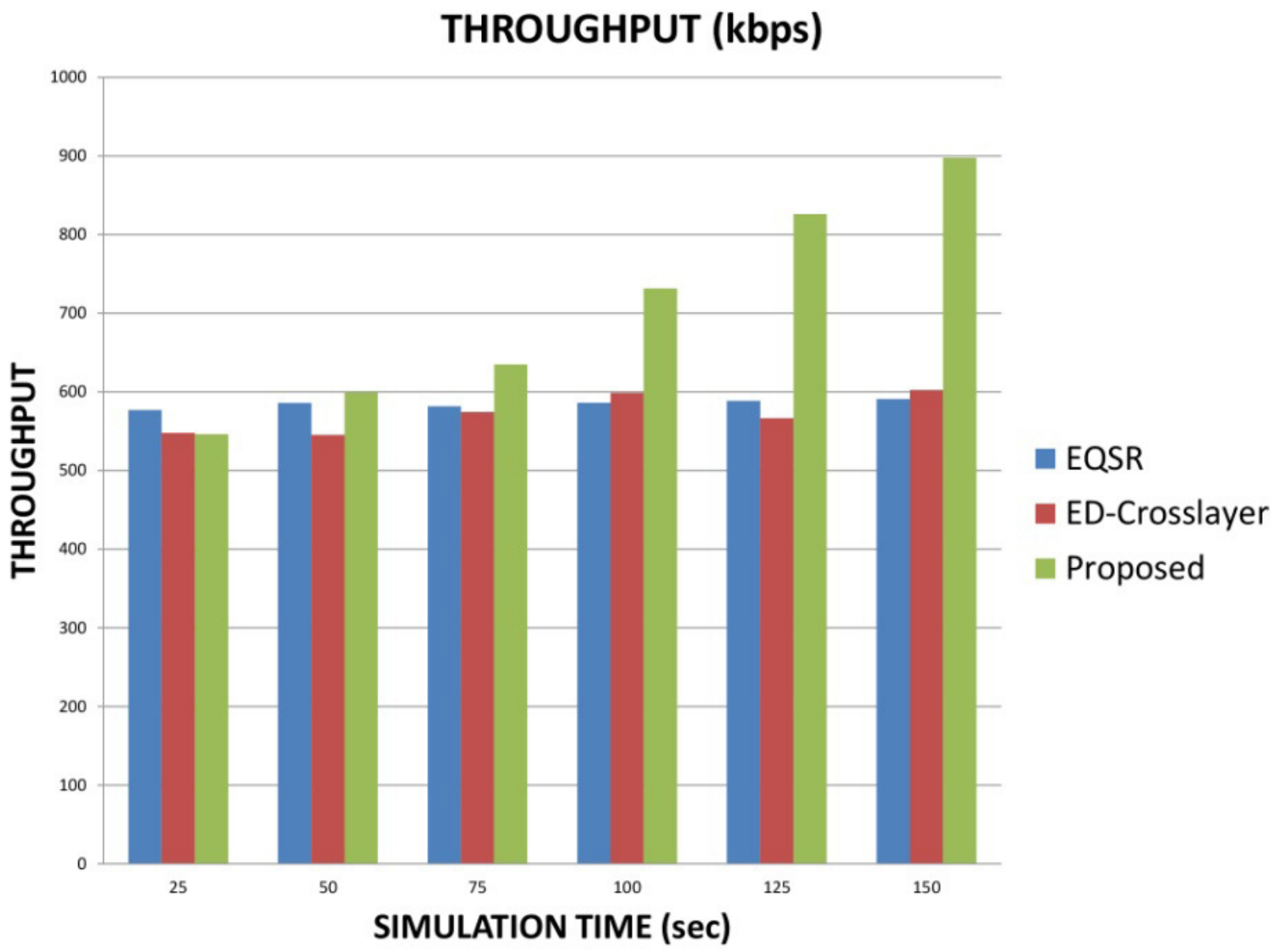

Figure 11 System Throughput Per Simulation Time

**End-to-End Delay**: This attains higher-levels in the modified approach when it is compared to offered load, as the offered load is increased then definitely it has an impact on the delay also, as shown in Figure (8), Whereas when it is compared with the no. of nodes, the end-to-end delay gets low as displayed in Figure (9).

**System Throughput**: This is expressed in kbps and is extremely prominent in the proposed approach when the offered load is increased which is shown in Figure (10), and also the throughput get increased even after comparing it with the simulation time shown in Figure (11). The throughput has efficient data rate transmission which is positively good when compared with the remaining conventional models.

## 8. CONCLUSION AND FUTURE SCOPE

A mobility error prediction & clustering multi-hop routing (CMHR) protocol have been implemented in this paper. The XLN approach was adopted as the best layer approach which is a combination of,

1. MAC/Data link layer whose purpose is to share the information of the energy levels, link capacity & received signal strength values which also acts as a service provider [16].
2. Network layer whose purpose is to establish the connection between the communication protocol like TCP (or) UDP [17].

Apart from the above two layers, transport layer also exists whose purpose is to control the communication by FTP (or) STTP (or) CBR [28]. This concept is having its feasibility as well as suitability for indoor applications [24]. By implementing the mobility error prediction model, the evaluation of remaining levels has been done. Also with the help of CH election & clustering multi-hop routing protocol, it was very feasible to choose the desired cluster head within the region to which it actually belongs [5, 18]. AODV as well as Two-Ray Ground Model has been preferred to provide the best results during the performance evaluation. Concluding that, it has been surveyed that the proposed model of the hybrid algorithm which is a combination of MEP as well as CHMHR outperforms better outputs when compared to the existing/conventional (EQSR, ED) models [4, 29] with respect to the variables such as energy consumption, end-to-end delay & throughput.

In the future, this method of approach can also be extended in terms of estimating other parameters like PDR, the number of live nodes, routing load with minimal usage of the energy consumption etc.

## 9. ABBREVIATIONS

The following abbreviations were used in this manuscript:

| | |
|---|---|
| XLN | Cross Layer Network |
| MEP | Mobility Error Prediction |
| CMHR | Cluster Multi-Hop Routing |
| MAC | Medium Access Control |
| PHY | Physical |
| N-layer | Network Layer |
| D/L layer | Data Link Layer |
| AODV | Adhoc On-Demand Distance Vector |
| DSR | Dynamic Source Routing |
| CSMA/CA | Carrier Sense Multiple Access/Collision Avoidance |
| LR-WAN's | Low Rate Wide Area Networks |
| DSDV | Destination-Sequenced Distance-Vector Routing |
| LLC | Logic Link Control |
| FFD | Full-Function Device |
| RFD | Reduced-Function Device |
| EQSR | Energy Efficiency & QoS Aware Multipath Routing |
| GCCR | Grid Based Clustering & Combinational Routing |
| PDR | Packet-to-Delivery Ratio |

## AUTHORS

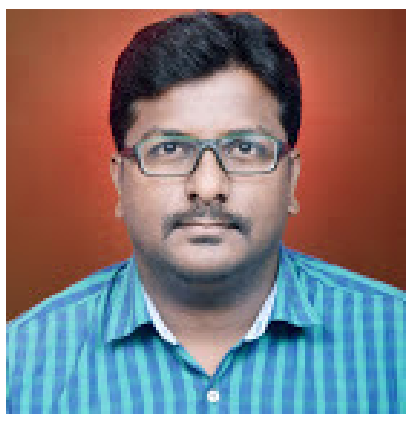

**Md. Khaja Mohiddin** received his B. Tech Degree in Electronics and Communication Engineering from Al-Ameer College of Engineering & Information Technology affiliated to JNTU, Andhra Pradesh, India, in 2009, M.Tech. Degree in Digital Electronics & Communication Systems from Chaitanya Engineering College affiliated to JNTU, Andhra Pradesh, India, in 2012. He is currently pursuing the Ph.D. in Wireless Sensor Network from GITAM University, Visakhapatnam, Andhra Pradesh, India. He is currently working as an Assistant Professor in the Department of Electronics and Telecommunication Engineering, Bhilai Institute of Technology, Raipur, (C.G.).

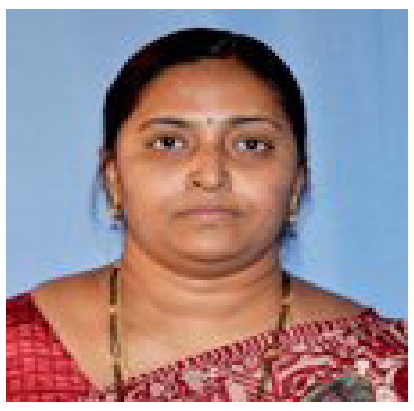

**V. B. S. Srilatha Indira Dutt** received her B. Tech. Degree in Electronics and Communication Engineering from Nagarjuna University, Andhra Pradesh, India, in 1994, M. Tech. Degree in Radar and Microwave Engineering from Andhra University, Andhra Pradesh, India, in 2007. She was awarded with Ph.D. degree in Global Positioning System from the Andhra University, Andhra Pradesh, India, in 2011. She is currently working as an Associate Professor in the Department of Electronics and Communication Engineering, GITAM Institute of Technology, GITAM University (GITAM), Visakhapatnam, Andhra Pradesh, India. Her research interests include Global Positioning System, Satellite Signal processing and Mobile Communications. She published more than 40 research papers in referred international journals, and international and national conferences.